\documentclass[mathleft]{an}
\usepackage{graphicx}
\usepackage{times}

\usepackage{natbib}
\usepackage{epstopdf}

\def\p{\partial}

\def\nab{\mbox{\boldmath $\nabla$}}

\def\rb{\bar{\rho}}

\def\tb{\bar{T}}
\def\sb{\bar{S}}

\newcommand{\bmv}{\mbox{\boldmath $v$}}
\newcommand{\DD}{\mbox{\boldmath ${\cal D}$}}

\begin{document}

\Pagespan{1}{}
\Yearpublication{2011}%
\Yearsubmission{2011}%
\Month{09}%
\Volume{999}%
\Issue{88}%

\title{Convection and Differential Rotation Properties of G \& K
  Stars Computed with the ASH Code}


\author{Sean P. Matt$^1$\thanks{Corresponding author:
  \email{sean.matt@cea.fr}\newline}
\and  Olivier Do Cao$^1$
\and  Benjamin P. Brown$^2$
\and  Allan Sacha Brun$^1$
}
\titlerunning{G \& K stars}
\authorrunning{Matt et al.}
\institute{
$^1$ Laboratoire AIM Paris-Saclay, CEA/Irfu Universit\'e Paris-Diderot
CNRS/INSU, 91191 Gif-sur-Yvette, France \\
$^2$ Department of Astronomy and Center for Magnetic Self Organization
in Laboratory and Astrophysical Plasmas, University of Wisconsin, 475 N. Charter
St., Madison WI 53706, USA}

\received{6 Oct 2011}
\accepted{17 Nov 2011}
\publonline{later}

\keywords{convection -- hydrodynamics -- Stars: interiors -- Stars: rotation -- turbulence}

\abstract{The stellar luminosity and depth of the convective envelope
  vary rapidly with mass for G- and K-type main sequence stars. In
  order to understand how these properties influence the convective
  turbulence, differential rotation, and meridional circulation, we
  have carried out 3D dynamical simulations of the interiors of
  rotating main sequence stars, using the anelastic spherical harmonic
  (ASH) code. The stars in our simulations have masses of 0.5, 0.7,
  0.9, and 1.1 $M_\odot$, corresponding to spectral types K7 through
  G0, and rotate at the same angular speed as the sun.  We identify
  several trends of convection zone properties with stellar mass,
  exhibited by the simulations.  The convective velocities,
  temperature contrast between up- and downflows, and meridional
  circulation velocities all increase with stellar luminosity.  As a
  consequence of the trend in convective velocity, the Rossby number
  (at a fixed rotation rate) increases and the convective turnover
  timescales decrease significantly with increasing stellar mass.  The
  3 lowest mass cases exhibit solar-like differential rotation, in a
  sense that they show a maximum rotation at the equator and minimum
  at higher latitudes, but the 1.1 $M_\odot$ case exhibits anti-solar
  rotation.  At low mass, the meridional circulation is multi-cellular
  and aligned with the rotation axis; as the mass increases, the
  circulation pattern tends toward a unicellular structure covering
  each hemisphere in the convection zone.}

\maketitle

\section{Introduction}


The sun and sun-like stars with convection zones in their outer
envelopes, have long been known to exhibit emission line and X-ray
activity, associated with hot gas in chromospheres, transition
regions, and coronae \citep[e.g.,][]{Strassmeier:1990p3592,
  Pizzolato:2003p3630, Wright:2011p3715}.  Understanding this activity, and
understanding how the solar case relates to the activity observed
accross the HR diagram is a long-standing puzzle.  It is clear that
the existence of hot gas above the photosphere is related to magnetic
processes associated with the convection zone itself.  The magnetic
field of low-mass main sequence stars is generally believed to be
generated by dynamo processes, which derive their properties from
convective motions, differential rotation, and meridional circulations
in the convection zones, as well as from the interaction between the
convection zone and the radiative interior.

Recent observations, by either spectropolarimetry
\citep[e.g.,][]{Donati:2003p3711}, doppler imaging
\citep[e.g.,][]{Barnes:2005p3591}, or monitoring of various activity
indicators \citep[e.g.,][]{Saar:1999p3589, Donahue:1996p3441,
  Baliunas:1995p3413, Lovis:2011p3719, Olah:2009p3742}, show that
solar-like stars possess activity cycles and differential rotation,
analogous to the sun.  Solar analogues are even starting to be
discovered \citep{Petit:2008p718}.

In order to gain a better theoretical understanding of how convective
properties depend upon stellar parameters, we carry out 3D numerical
dynamical simulations of convective envelopes of solar-like stars.  As
a first step, we model here the convective regions only, neglecting
the effects of the interface region with a radiative envelope, and
restrict ourselves to relatively slow (solar) rotation rates.
Specifically, we simulate the convection dynamics for 4 main sequence
stars with masses of 0.5, 0.7, 0.9, and 1.1 solar masses, spanning
spectral types G0 to K7.  This mass interval exhibits a large range of
the physical properties of convective envelopes (such as the depth,
physical size, mass, and density), as well as in the overall stellar
luminosity transported by convection.  These types of stars are also
targets for asteroseismic studies \citep[e.g.,][]{Verner:2011p3761},
which have the potential to give precise measurements of stellar
properties for large numbers of stars.  The goal here is to determine
how the convection, differential rotation, and meridional circulation
is influenced by stellar mass, and to see if general trends or scaling
laws can be extracted that will guide a deeper understanding of the
inner hydrodynamics of these stars.  The present study lays the
groundwork for later studies to consider (e.g.) faster rotation rates,
convection-radiation zone interface dynamics, and the dynamo
generation of magnetic fields in stars in this mass range.

Section \ref{sec_method} contains a description of our simulation
method and presents a comparison of the overall structures of each
star.  Section \ref{sec_results} describes the main results of our 3D
simulations, focusing on both the convective turbulence properties, as
well as the differential rotation and meridional circulation flows.  A
summary and brief discussion is contained in section
\ref{sec_summary}.

\section{Simulation method}
\label{sec_method}

We use the anelastic spherical harmonic (ASH) code
\citep[][]{clunea99} to compute the 3-dimensional and turbulent flows
in convectively unstable stellar envelopes.  This code has been
extensively tested and used for computing several aspects of the solar
interior \citep[e.g.,][]{Brun:2002p1, DeRosa:2002p3258, Brun:2004p6,
  Miesch:2006p1251, Browning:2006p1354}, rapidly rotating young stars
\citep{Ballot:2007p1132, Brown:2008p1119, Brown:2009p1178,
  Brown:2011p1815}, the convective cores of massive stars
\citep{Browning:2004p3582, Browning:2004p3583,
  Featherstone:2009p1462}, fully convective low mass stars
\citep{Browning:2008p1139}, red giant stars \citep{Brun:2009p3585},
and pre-main-sequence stars \citep[Bessolaz \& Brun, in this volume;
][]{Bessolaz:2011p3250}.  We briefly describe the basic aspects of the
code here, but the reader can find further details of the code in
those previous works \citep[see especially,][]{clunea99, Brun:2004p6}.

The code solves the fluid equations, under the anelastic
approximation, in a computational domain consisting of a spherical
shell and in a rotating reference frame.  Under the anelastic
approximation, sound waves are filtered out and assumed to have a
negligible effect on the dynamics, in order to allow for a larger
computational timestep.  This approximation is appropriate (e.g.) in
the interiors of stars because typical motions are highly sub-sonic.  The
thermodynamic variables are linearized with respect to a spherically
symmetric background state with a density $\rb$, pressure $\bar P$,
temperature $\tb$, and specific entropy $\sb$ and fluctuations about
the background state of $\rho$, $P$, $T$, and $S$.  The time-dependent
equations describe the conservation of mass, momentum, and entropy
expressed as
\begin{equation}
\label{eq_mass}
\nab\cdot(\rb \bmv) = 0,
\end{equation}
\begin{eqnarray}
\label{eq_momentum}
\rb \left(\frac{\p \bmv}{\p t}+(\bmv\cdot\nab) \bmv +  2{\bf
\Omega_o}\times \bmv\right) =
\;\;\;\;\;\;\;\;\;\;\;\;\;\;\;  \nonumber \\
-\nab P + \rho {\bf g}
- [\nab\bar{P}-\rb{\bf g}] 
- \nab\cdot\mbox{\boldmath $\cal D$},
\end{eqnarray}
\begin{eqnarray}
\label{eq_entropy}
\rb &\tb& \frac{\p S}{\p t} + \rb \tb \bmv\cdot\nab (\sb+S) = \nonumber \\ 
&+& \nab\cdot[\kappa_r \rb c_p \nab (\tb+T) +
  \rb \tb (\kappa \nab S + \kappa_0 \nab \sb)] \nonumber \\ 
&+& 2\rb\nu\left[e_{ij}e_{ij}-1/3(\nab\cdot \bmv)^2\right],
\end{eqnarray}
where $\bmv = (v_r, v_\theta, v_\phi)$ is the velocity in the rotating
frame in spherical coordinates, ${\bf \Omega_0} = \Omega_0 {\bf \hat
  e_z}$ is the angular rotation rate of the reference frame, ${\bf g}$
is the acceleration due to gravity, $\kappa_r$ is the radiative
diffusivity, and $c_p$ is the specific heat at constant pressure.  The
term $\DD$ is the viscous stress tensor, with the components
\begin{eqnarray}
\label{eq_viscoustensor}
{\cal D}_{ij}=-2\rb\nu[e_{ij}-1/3(\nab\cdot \bmv)\delta_{ij}],
\end{eqnarray}
where $e_{ij}$ is the strain rate tensor, and $\nu$, $\kappa$, and
$\kappa_0$ are effective eddy diffusivities.  The code also uses a
linerized equation of state,
\begin{equation}
\label{eq_eos}
\frac{\rho}{\rb}=\frac{P}{\bar{P}}-\frac{T}{\tb}=\frac{P}{\gamma\bar{P}}
-\frac{S}{c_p},
\end{equation}
and the ideal gas law,
\begin{equation}
\label{eq_idealgas}
\bar{P}={\cal R} \rb \tb ,
\end{equation}
where $\gamma$ is the ratio of specific heats (we use $\gamma=5/3$),
and ${\cal R}$ is the gas constant.  For all 4 stars in our study, the
luminosity has reached a value of more than 99.9\% of the total
stellar luminosity by the base of the convection zone, so we do not
need to include any energy generation by nuclear burning within our
computational domain.


We set up four different models in which the domain boundaries and
stratification coincides with 1D models of the convection zones of main
sequence stars with different masses.  Section \ref{sec_1d} describes
the global properties of these stars, and sections \ref{sec_init} and
\ref{sec_ss} describe the initial and boundary conditions used in our
ASH models, as well as the method for evolving the simulations to a
fully-convective, statistical steady-state.

\subsection{1D stellar structure} \label{sec_1d}

\begin{table*}
\centering
  \caption{Global properties of the 4 stars used in our ASH models, 
    computed with the CESAM
    stellar evolution code and at an age of 4.6 Gyr.  We adopt $M_\odot =
    1.989 \times 10^{33}$g, $R_\odot = 6.9599 \times 10^{10}$ cm, 
    and $L_\odot = 3.846 \times 10^{33}$ erg s$^{-1}$.}
\label{tab_properties}
\begin{tabular}{cccccccccccccc}
\hline

Mass & Radius & $L_*$ & $T_{\rm eff}$ & SpT & $M_{\rm cz}$ & $R_{\rm
 cz}$ & $T(R_{\rm cz})$ & $\rho(R_{\rm cz})$ \\ 

($M_\odot$) & ($R_\odot$) & ($L_\odot$) & (K) &  &
($M_\odot$, $M_*$) & ($R_\odot$, $R_*$) & (K) & (g cm$^{-3}$) \\

\hline

0.5 & 0.44 & 0.046 & 4030 & K7       & 0.18, 0.36     & 0.25, 0.56 &
          $4.3\times10^6$ & 14 \\  
0.7 & 0.64 & 0.15   & 4500 & K4/K5 & 0.079, 0.11    & 0.42, 0.66 &
          $3.0\times10^6$ & 2.1 \\  
0.9 & 0.85 & 0.55   & 5390 & G8      & 0.042, 0.046   & 0.59, 0.69 &
          $2.6\times10^6$ & 0.51 \\  
1.1 & 1.23 & 1.79   & 6030 & G0      & 0.011, 0.0100 & 0.92, 0.75 &
          $1.6\times10^6$ & 0.048 \\ 

\hline

\end{tabular}
\end{table*}

\begin{figure}
\includegraphics[width=\linewidth]{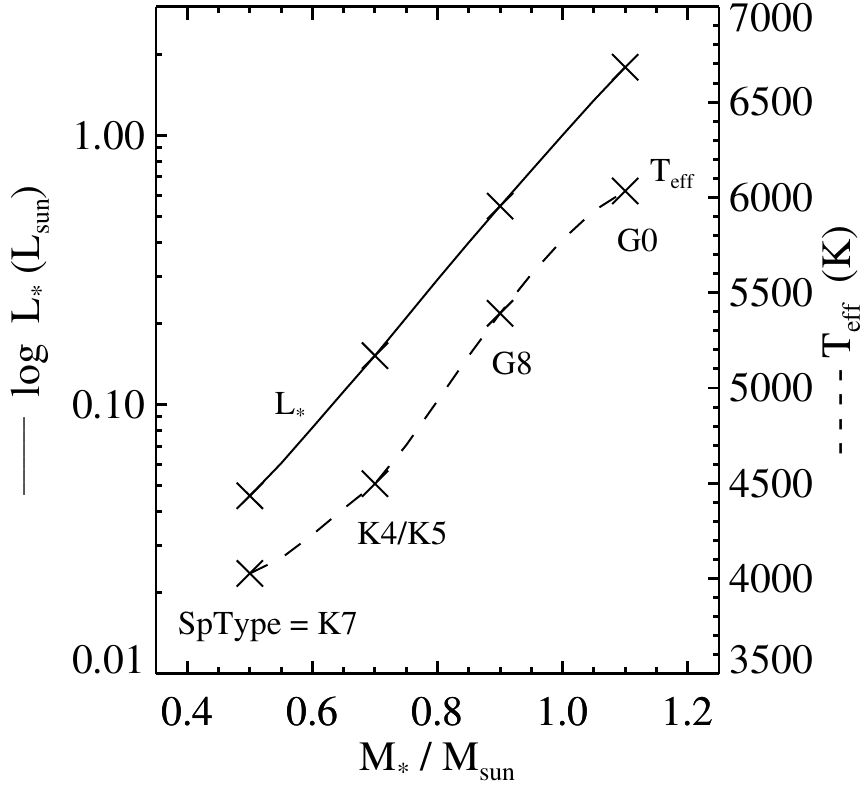}
\caption{Stellar luminosity in solar units (solid line, left scale)
  and effective temperature (dashed line, right scale) as a function
  of stellar mass, for the mass range spanned by our models, and as
  computed using the CESAM stellar evolution code.  The ``X'' symbols
  indicate the values for the 4 stars used in our 3D ASH simulations.
  The approximate spectral types are also indicated next to each of
  these.}
\label{fig_lt}
\end{figure}

\begin{figure}
\includegraphics[width=\linewidth]{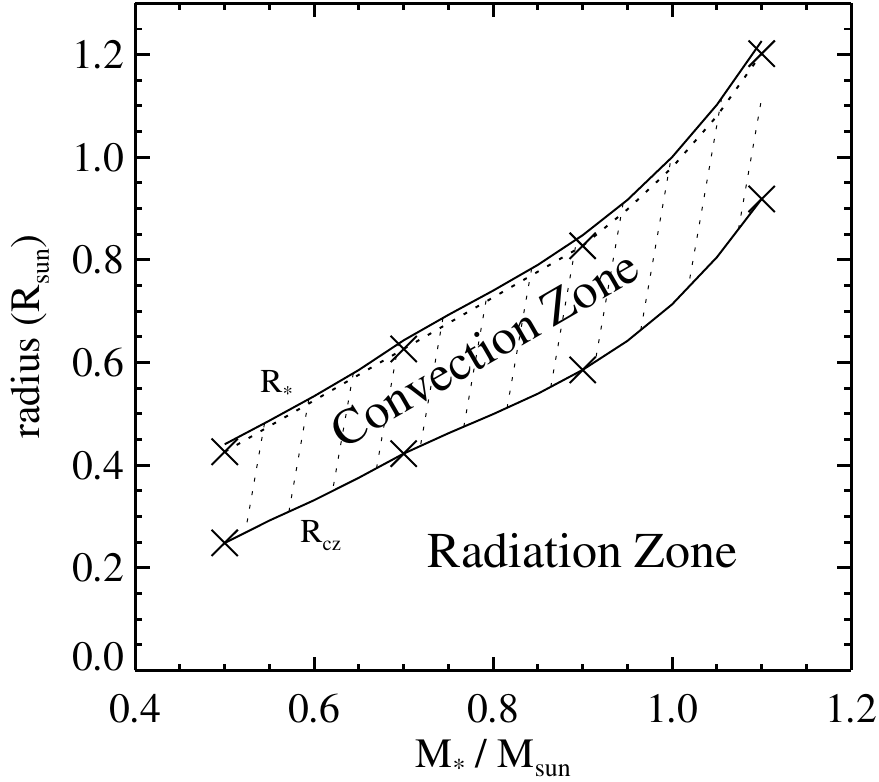}
\caption{Stellar radius (upper line) and radial location of the base
  of the convection zone (lower line) in solar units and as a function
  of stellar mass, as computed by the CESAM stellar evolution code.
  The convection zone exists between these two lines, while a
  convectively stable radiation zone lies below.  The ``X'' symbols
  indicate the upper and lower radial boundaries of the computational
  domain used in our ASH simulations, for the 4 cases considered.}
\label{fig_rcz}
\end{figure}

\begin{figure}
\includegraphics[width=\linewidth]{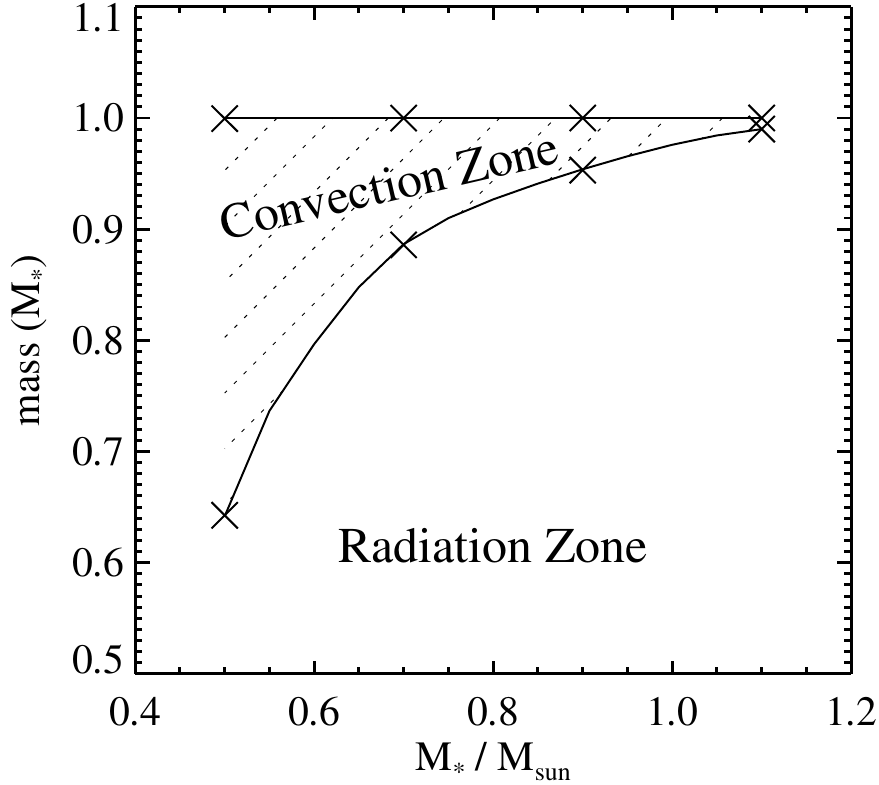}
\caption{Mass contained beneath the stellar surface (upper line) and
  location of the base of the convection zone (lower line) as a
  function of stellar mass, as computed by the CESAM stellar evolution
  code.  The convection zone exists between these two lines, while a
  convectively stable radiation zone lies below.  The ``X'' symbols
  indicate the upper and lower radial boundaries of the computational
  domain used in our ASH simulations, for the 4 cases considered.}
\label{fig_mcz}
\end{figure}

In order to define the background structure for our 3D models, we use
the 1D stellar evolution code CESAM \citep[][]{Morel:1997p3487}.  With
CESAM, we computed the evolution of four stars with masses of 0.5,
0.7, 0.9, and 1.1 $M_\odot$ until the age of 4.6 Gyr.  This age is
approximately equal to that of the sun, so our cases can be compared
with the many previous results of solar studies.  Also, this age is
appropriate for the slow (solar) rotation rates considered here.  For
all four stars, we assumed the same initial metallicity of (X, Y, Z)
$\approx$ (0.71, 0.27, 0.02) and a mixing length parameter of 1.77,
chosen to best represent the solar case.  Table \ref{tab_properties}
lists the key global properties of our four stars, at the age of 4.6
Gyr, which are also graphically represented in Figures
\ref{fig_lt}--\ref{fig_mcz}.

Figure \ref{fig_lt} shows the total luminosity (solid line), as well
as the effective temperature (dashed line) as a function of mass, for
stars in the mass range of our models.  The ``X'' symbols mark the
values for the 4 stars modeled here, which are also listed in table
\ref{tab_properties}.  Also indicated on the plot are the approximate
spectral types, corresponding to the effective temperature.  The
Figure demonstrates the steep rise in lumonosity with mass expected
for main sequence stars (the stars here approximately follow $L_*
\propto M_*^{4.6}$), such that the 1.1 $M_\odot$ star is 40 times more
luminous than the 0.5 $M_\odot$ star.

Figure \ref{fig_rcz} shows the photospheric radius $R_*$ (upper solid
line) and the radial location of the base of the convection zone
$R_{\rm cz}$ (lower solid line) as a function of mass.  The stellar
radius is strong function of stellar mass in the plotted range, with
the 1.1 $M_\odot$ star being 2.8 times bigger than the 0.5 $M_\odot$
star.  The radial extent of the convection zones for these stars is
represented by the region between the two solid lines, and the region
below contains a convectively stable radiation zone.  It is clear
that, in the mass range shown, the thickness of the convection zone
increases slightly with increasing stellar mass.  However, since the
stellar radius increases more rapidly, the fractional size of the
convection zone decreases with increasing mass.  Thus, the convection
zone thickness ranges from 44\% to 25\% of the stellar radius, for the
0.5 and 1.1 $M_\odot$ stars, respectively (for the sun, this value is
approximately 30\%).  The ``X'' symbols indicate the upper and lower
boundaries of our 3D simulation domains for the 4 stars simulated
(discussed below).

\begin{table*}
\centering
  \caption{Simulation parameters for each case.  The nuber of radial,
    latitudinal, and longitudinal gridpoints are
    $N_r$, $N_\theta$, and $N_\phi$.  The outermost radius of the
    simulated spherical shell is $R_{\rm out}$, and $H_\rho(R_{\rm
      out})$ is the density scale height there.  The radial size of
    the domain is $L=R_{\rm out}-R_{\rm cz}$.  The viscosity at
    $R_{\rm out}$ is $\nu_{\rm top}$ and it varies in the domain with the inverse
    square root of the background density.  All stars have a
    Prandtl number $P_r \equiv \nu/\kappa = 0.25$ and rotate at the
    solar rate, $\Omega_0 = 2.6 \times 10^{-6}$ rad/s, corresponding to a
    rotation period of $2 \pi / \Omega_0 \approx 28$ days. Also listed are
    the Rayleigh number $R_a \equiv (-\partial \rho / \partial S) \Delta
    S g L^3 / (\rho \nu \kappa)$, the Taylor number $T_a \equiv 4
    \Omega^2 L^4/\nu^2$, and the convective Rossby number $R_{oc} \equiv
    (R_a/T_a P_r)^{1/2}$, all evaluated at the midlevel of
    the domain.}
\label{tab_parameters}
\begin{tabular}{cccccccccccccc}
\hline

Mass &  $N_r, N_\theta, N_\phi$ &  $R_{\rm out}$ & $H_\rho(R_{\rm out})$ &
  $L$ & $\nu_{\rm top}$ & $R_a$ & $T_a$ & $R_{oc}$ \\

($M_\odot$) &  & ($R_\odot$, $R_*$)  & (Mm) & (Mm) & (cm$^2$ s$^{-1}$) &&& \\

\hline  

0.5 & 257, 256, 512   & 0.43, 0.97 & 6.1 & $120$ &
          $3.5\times10^{11}$ & $3.9\times10^{6}$ & $1.5\times10^{8}$ & 0.32 \\  
0.7 & 257, 256, 512 & 0.63, 0.97 & 6.3 & $140$ &
          $2.0\times10^{12}$ & $6.0\times10^{5}$ & $8.1\times10^{6}$ & 0.54 \\  
0.9 & 257, 256, 512 & 0.83, 0.98 & 7.6 & $170$ &
          $5.0\times10^{12}$ & $1.9\times10^{5}$ & $2.6\times10^{6}$ & 0.54  \\  
1.1 & 257, 256, 512   & 1.20, 0.98 & 8.6 & $200$ &
          $2.0\times10^{13}$ & $3.6\times10^{4}$ & $3.2\times10^{5}$ & 0.67 \\

\hline
\end{tabular}
\end{table*}

Figure \ref{fig_mcz} shows the extent of convection zones, expressed
in mass coordinates and normalized to the mass of each star, as a
function of stellar mass.  The lines show the total mass enclosed by
the stellar surface (upper line) and enclosed by the location of the
base of the convective envelope (lower line).  The amount of mass
contained in the convection zone, $M_{\rm cz}$, is the difference
between these two lines and is listed in Table \ref{tab_properties}.
It is clear that the convection zone mass is a strong function of
stellar mass in the range considered.  Stars with slightly lower mass
will be fully convective, while significantly more massive stars will
not have convective envelopes.  The convection zone mass varies from
36\% to 1\% of the stellar mass, for the 0.5 and 1.1 $M_\odot$ stars,
respectively (the solar value is approximately 3\%).  The ``X''
symbols indicate the simulation domain boundaries, expressed in mass
coordinates, for the 4 stars simulated (discussed below).

The last two columns of table \ref{tab_properties} show the
temperature and the mass density at the base of the convection zone
for each case.  Although it is true that the temperature and density
increases with mass at the stellar center ($R=0$), these quantities
decrease with increasing mass at the location of $R_{\rm
  cz}$.  In other words, for increasing stellar mass, the convection
zone generally occupies a more tenuous and cooler outer layer of the
star.

\subsection{Initial and boundary conditions for 3D simulations}
\label{sec_init}

We use the 1D stellar structure models described above for the
spherically symmetric, initial conditions of the the 3D ASH models.
Table \ref{tab_parameters} lists the key properties of each of the 4
simulations.  In the ASH models presented here, the domain consists of
the stellar convection zone only.  Thus, the lower radial boundary of
our simulated spherical shells coincides with the base of the
convection zone, $R_{\rm cz}$, for each star (see table
\ref{tab_properties} and the lower row of ``X'' symbols in Figs.\
\ref{fig_rcz} and \ref{fig_mcz}).  We chose the location of the outer
boundary such that the mass density there is a factor of 100 smaller
than at the base of the convection zone.  This places the outer
boundary somewhat below the photosphere of the star.  The location of
the outer boundary $R_{\rm out}$ is listed in table
\ref{tab_parameters} and shown as the upper row of ``X'' symbols in
and Figures \ref{fig_rcz} and \ref{fig_mcz}.

In each of our 4 cases, the gravity ($\mbox{\boldmath $g$}$) is taken
directly from the corresponding CESAM model, and the entropy gradient
($\partial \sb/\partial r$) is initialized with a constant value equal
the (superadiabatic) entropy gradient in the middle of the convection
zone in the CESAM model.  The background thermodynamic variables
($\rb$, $\bar P$, $\tb$, and $\sb$) are set according to this entropy
gradient and to be in hydrostatic balance with the stellar gravity,
while the fluctuating thermodynamic variables ($\rho$, $P$, $T$, and
$S$) are initially zero.  The initial entropy gradient and choice of
diffusivities (see below) ensures that the initial state is
convectively unstable and has a Rayleigh number near or above the
critical Rayleigh number necessary for convection.  The velocity in
the rotating frame ($\bmv$) is given an initial random perturbation,
with negligible kinetic energy, in order to initiate convective
motions.

For both the inner and outer boundaries, we use stress-free and
impenetrable boundary conditions on the velocity and hold the entropy
gradient fixed at the initial value.  The radiative diffusivity
$\kappa_r$ is chosen so that the ASH model has the same radiative flux
at all radii as the corresponding CESAM model.  The initial stellar
structure and boundary conditions ensure that the energy flux into the
domain at the bottom of the convection zone (given entirely by the
radiative flux) is constant in time.  At the top of the domain, the
density scale height becomes small (see table \ref{tab_parameters}),
and since this approximately determines the convection cell size
scale, it becomes numerically challenging to resolve the convective
motions there.  Furthermore, the impenetrable boundary condition
precludes any convective enthalpy flux from escaping the top of the
domain.  To address both of these issues we introduce a diffusive
energy flux (the term proportional to $\kappa_0$ in eq.\
[\ref{eq_entropy}]) that is assumed to represent an unresolved and
spherically symmetric enthalpy flux carried by small-scale convection
near the top of the domain (hereafter ``unresolved eddy flux'').  In
all 4 cases, we chose $\kappa_0$ so that this flux is negligible in
the bulk of the convection zone, but increases as a smooth function
near the outer boundary, such that the flux leaving the domain equals
the stellar energy flux and is constant in time.

\begin{figure*}
\includegraphics[width=0.485\linewidth]{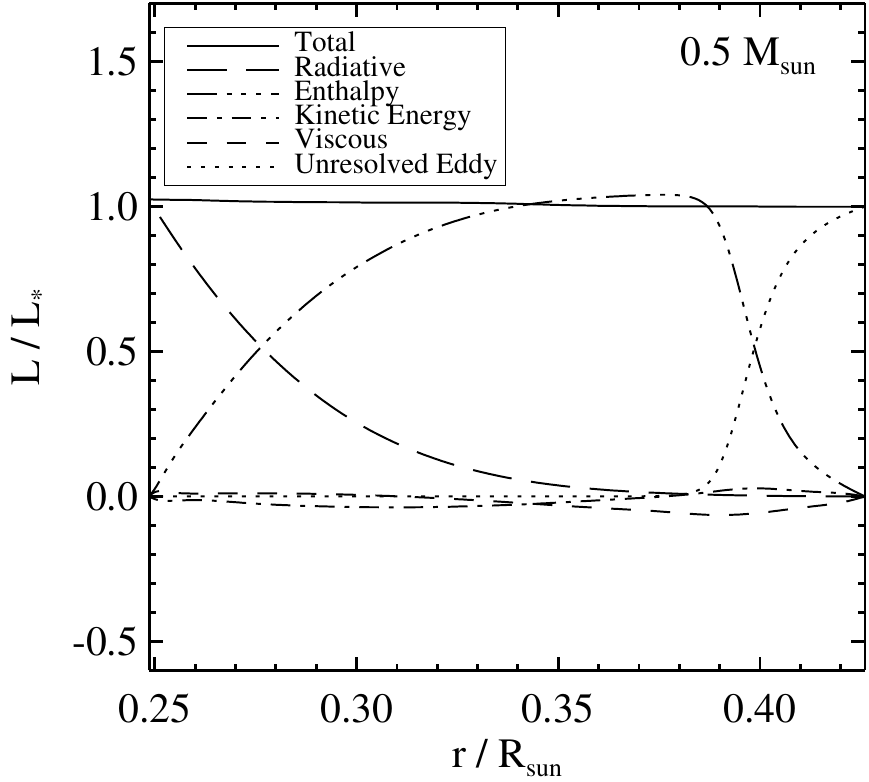}
\includegraphics[width=0.5\linewidth]{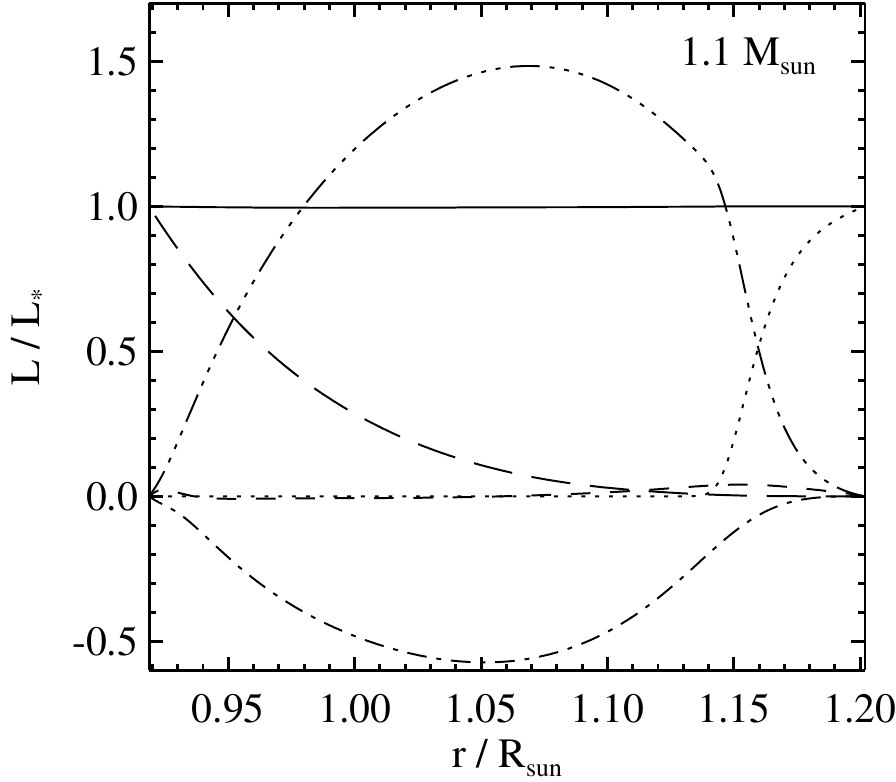}
\caption{Energy flow in 0.5 (left) and 1.1 $M_\odot$ (right) cases as
a function of radius in the computational domain.  Shown are the
spherically- and time-averaged luminosities, after the simulations
have reached a statistical steady-state.}
\label{fig_fluxbalance}
\end{figure*}

In each case, we chose the viscosity at the outer part of the domain,
$\nu_{\rm top}$, in order to ensure that the resulting convective
turbulence has a significant Reynolds number.  As shown in section
\ref{sec_results}, a higher stellar luminosity leads to larger
convective velocities, and thus a larger viscosity is needed to keep
the Reynolds numbers comparable (explaining the trend of $\nu_{\rm
  top}$ apparent in table \ref{tab_parameters}).  Within the domain,
the viscosity in each case depends only upon radius, and it varries as
the inverse square root of the background density.  Scaling the
viscosity in this way ensures a significant turbulence level at all
radii and is used in most previous ASH models in the literature.  The
thermal diffusivity $\kappa$ equals $4\nu$ everywhere in the domain,
giving a constant Prandtl number of 0.25 for all cases.  The rotation
rate of the reference frame ($\Omega_0$), which equals the rotation
rate of the star, is set to the solar rate for all 4 stars.

The Rayleigh numbers (calculated using the steady-state value of the
entropy in the simulations), Taylor numbers, and convective Rossby
numbers for each case are also given in table \ref{tab_parameters}.
The Rayleigh and Taylor numbers, show a strong decrease with
increasing stellar mass, which is primarily explained by the trend in
the diffusivities.  If the diffusivities were constant, these numbers
would increase with mass for these stars.  The convective Rossby
number generally increases with mass, meaning that for the same
rotation rate, the coriolis force has less influence on the dynamics.

\subsection{Reaching a statistical steady-state and energy flux balance}
\label{sec_ss}

At the start of the simulation, the star is in a quiescent state.
Thus, there is no significant enthalpy flux, and the system is not
initially in radial energy flux balance.  In this situation, the
evolution of the system, according to equations
(\ref{eq_mass})--(\ref{eq_entropy}), ensures that the entropy gradient
evolves toward an energy flux balanced state.  Furthermore, since the
gas in the computational domain is unstably stratified ($\partial \sb
/ \partial r < 0$), with a large Rayleigh number (see table
\ref{tab_parameters}), significant convective motions begin rapidly
after the start of the simulations.  Once vigorous convection begins,
an energy flux balance is achieved within a few convective turnover
times (months, typically) and maintained for the duration of the
simulations.  In this state, the net energy transport across the
domain is constant at all radii, when averaged over several convective
turnover times.

Figure \ref{fig_fluxbalance} shows the spherically- and time-averaged
luminosity, as a function of radius in the whole domain, for the
highest and lowest mass stars in this study.  These are shown after
the simulated stars have evolved for several years from the
initial state.  The boundary conditions ensure a fixed radiative
energy flux into the domain at the inner boundary and a fixed
unresolved eddy flux out of the domain at the outer boundary, both
with luminosities equal to that of the modeled stellar luminosity.
Within the domain, the radiative energy flux is nonzero, but it
decreases in importance with increasing radius in the convection zone.
The enthalpy flux, associated with convective motions, carries the
bulk of the remaining energy flux from the lower boundary, across the
convection zone, to the outer boundary.  Due to the asymmetry between
broad, slow upflows and narrow, fast downflows, there is generally a
net negative (downward) flux of kinetic energy in the convection zone
(e.g., as in the right panel of Fig.\ \ref{fig_fluxbalance};
\citealp{Cattaneo:1991p3746, Hurlburt:1986p3743}).  As a result, the
steady-state enthalpy luminosity (dash-triple-dotted line) peaks at a
value significantly larger than $L_*$, in order that the net energy
flow across the convection zone is constant and equal to the stellar
luminosity.  As demonstrated by the two extreme cases in Figure
\ref{fig_fluxbalance}, the presence of a super-luminal enthalpy flux
is increasingly important with mass, in our 4 simulations.  It is not
clear whether this is due to an intrinsic property of the stars (such
as the luminosity), or whether it only depends upon (e.g.) the
viscosity, which is systematically different in each of our cases
(table \ref{tab_parameters}).  However, it does seem that the faster
flows found in the more massive stars (discussed below) yields a
larger negative kinetic energy flux, since the difference between the
broad-slow upflows and narrow-fast downflows is accentuated.

\begin{figure}
\includegraphics[width=\linewidth]{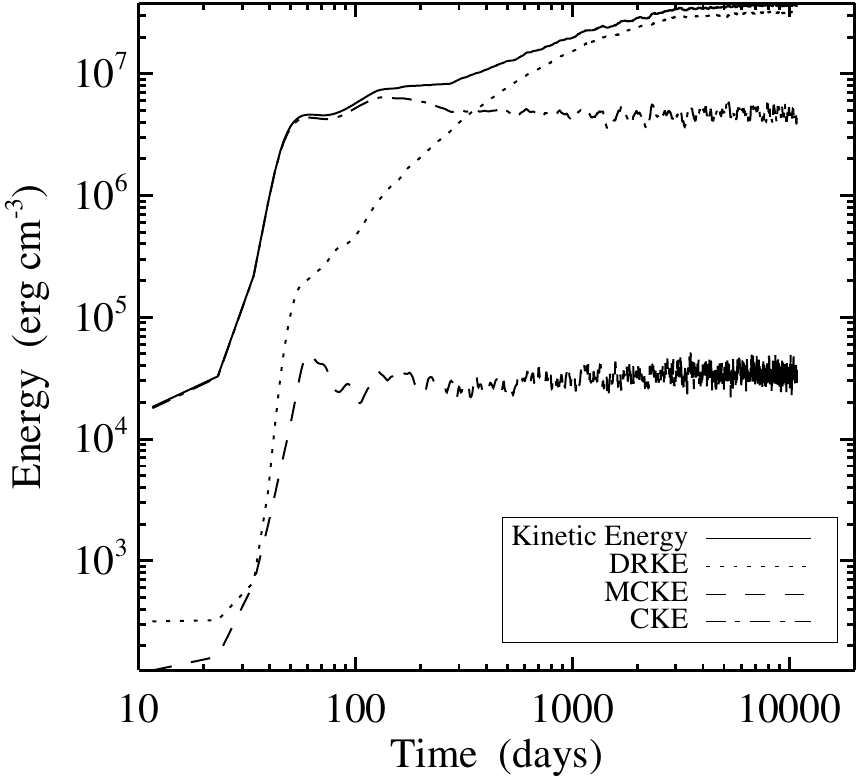}
\caption{Integrated kinetic energies in the domain, divided by the
  total volume, as a function of time, for the 0.5 $M_\odot$ case.
  Shown are the kinetic energies associated with the differential
  rotation (DRKE), meridional circulation (MCKE), convective motions
  (CKE), and the total (solid line).}
\label{fig_scalareng}
\end{figure}

Although the energy flux balance is established relatively rapidly in
the convection zone, the statistical properties of the turbulence can
take longer to settle down, as they evolve in response to (e.g.)
Reynolds stresses and the dissipation of energy on small scales.  In
addition, these systems typically exhibit global, axisymmetric flows
such as differential rotation and meridional circulation.  These flows
are also influenced by the Reynolds stresses, and the meridional
circulation typically has velocities smaller than the convective
velocities.  Thus, the global flows settle down on timescales of the
order of 10 times the convective turnover time.  In order to ensure
the simulations have time to establish these global flows and are near
their turbulent statistical steady-state, we have run each model for a
timescale comparable to the global viscous dissipation timescale
($L^2/\nu$).

\begin{table*}
\centering
\caption{Simulation results for each case.  The amount of evolution
  time since the start of the simulation is denoted $t_f$.  Temporal averages of the
  rms components of velocity $\tilde v_r$, $\tilde v_\phi$, $\tilde v_\theta$,
  speed $\tilde v$, and fluctuating velocities $\tilde v^\prime_\phi$ and
  $\tilde v^\prime$ are evaluated at the midlevel of the domain and
  given in units of m s$^{-1}$.  The temporally averaged rms
  temperature fluctuation at midlevel is $\tilde T$.  Also
  listed are the rms Reynolds number $\tilde R_e \equiv \tilde v^\prime L/\nu$,
  Rossby number $\tilde R_o \equiv \tilde v^\prime /(2 \Omega_0 L)$, and
  P\'eclet number $\tilde P_e \equiv \tilde v^\prime L/\kappa$, evaluated at
  the midlevel, and the convective turnover timescale $\tau_{\rm to} \equiv
  L / \tilde v_r$.  The $\Delta \Omega$ is the difference in angular
  rotation rate at the outer
  boundary, between latitudes of 0$^\circ$ and 60$^\circ$.  The
  $\Delta T$ is the difference in the temporally and azimuthally
  averaged temperature at the base of the convection zone ($R_{\rm
    cz}$), between latitudes of 60$^\circ$ and 0$^\circ$.  Finally,
  $\tilde v_{\rm mc}$ is the rms meridional circulation speed,
  calculated from temporally and azimuthally averaged poloidal
  velocity and by taking the rms of all values at a constant midlevel radius.}
\label{tab_results}
\begin{tabular}{cccccccccccccccc}
\hline

Mass & $t_f$ & $\tilde v_r$ & $\tilde v_\theta$ & $\tilde v_\phi$ & 
             $\tilde v^\prime_\phi$ & $\tilde v$ & $\tilde v^\prime$ & 
             $\tilde T$ & $\tilde R_e$ & $\tilde R_o$ & $\tilde P_e$ &
            $\tau_{\rm to}$ & $\Delta\Omega / \Omega_0$ & 
            $\Delta T$ & $\tilde v_{\rm mc}$  \\ 

($M_\odot$) & (years) & (m/s) &(m/s) & (m/s) & (m/s) & (m/s) & (m/s) &
(K) & & &  &
(days) & (\%) & (K) & (m/s)  \\

\hline

0.5 & 30 & 8.7  & 9.8   & 37   & 9.3  & 40   & 16   & 0.22 & 309 & 0.025 & 
        77 & 164 & 37 & 0.61 & 0.29   \\  
0.7 & 28  & 20   & 21   & 74   & 21   & 79   & 36   & 0.76 & 139 & 0.049 & 
         35 & 82 & 41 & 2.2 & 0.64  \\  
0.9 & 22  & 47   & 45   & 95   & 50   & 115 & 82   & 1.3   & 151 & 0.094 & 
         38   & 41  & 30 & 3.6 & 1.5  \\   
1.1 & 8.6  & 150 & 134 & 242 & 136 & 314 & 241 & 6.1   & 133 & 0.24   & 
         33   & 15 & $-42$ & $-9.9$ & 22  \\  

\hline
\end{tabular}
\end{table*}

\begin{figure*}
\centering
\includegraphics[width=.99\linewidth]{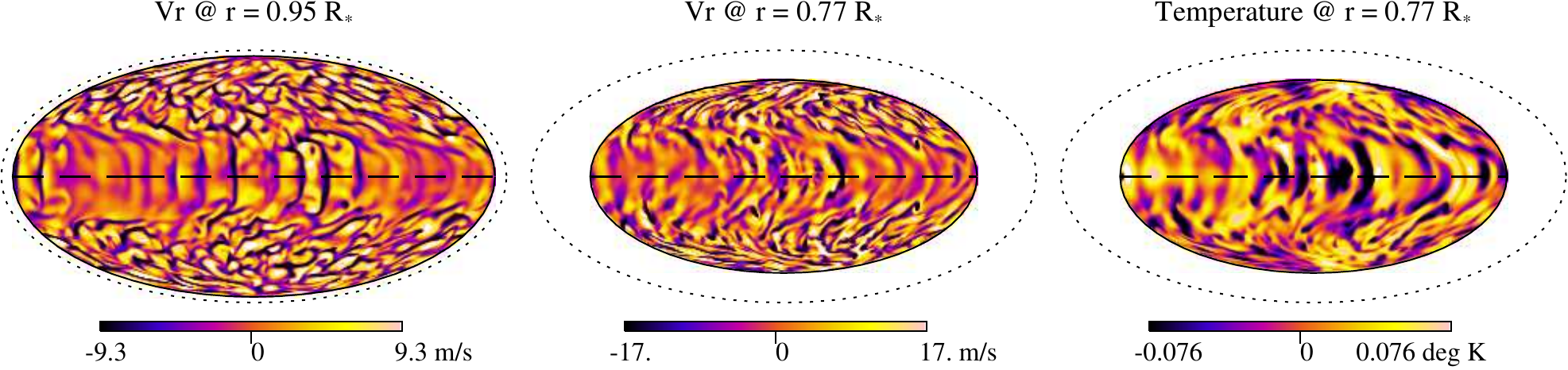}
\includegraphics[width=.99\linewidth]{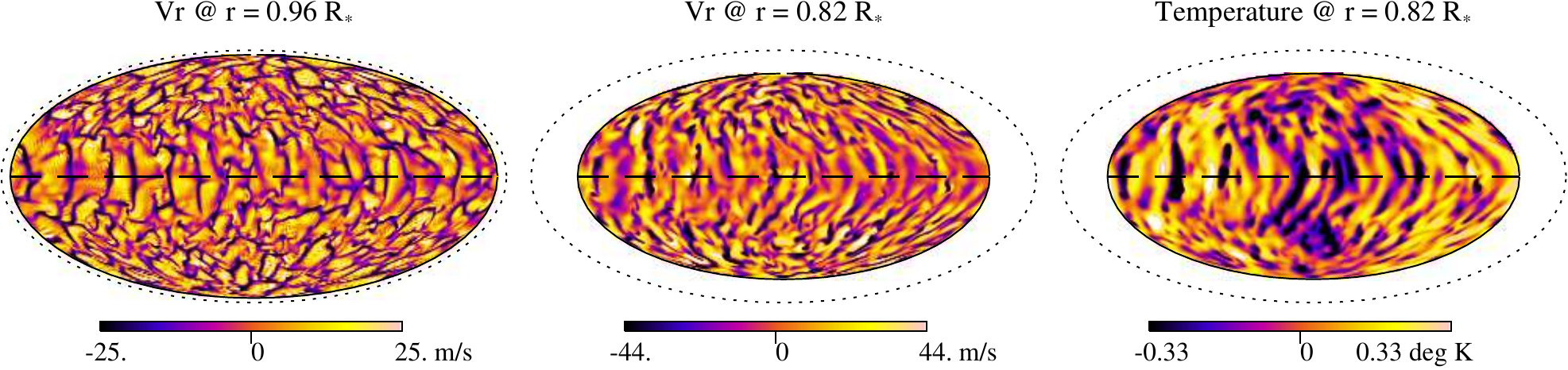}
\includegraphics[width=.99\linewidth]{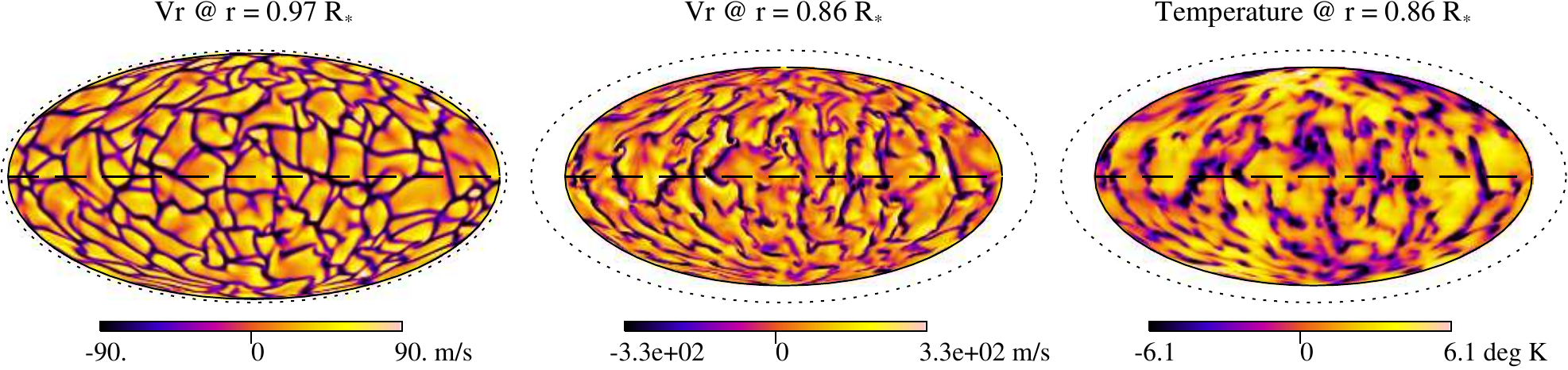}
\caption{Slices at constant radius (Mollweide view) for the 0.5
  $M_\odot$ case (top), 0.7 $M_\odot$ case (middle), and 1.1 $M_\odot$
  case (bottom), showing the radial velocity near the surface (left
  panels) and in the midlevel (middle panels) of the domain and the
  temperature at midlevel (right panels).  In the temperature slices,
  the azimuthal average has been subtracted, to emphasize the
  temperature fluctuations.  The dotted line corresponds to the
  stellar surface and the dashed line to the equator.}
\label{fig_ss}
\end{figure*}

As an example, Figure \ref{fig_scalareng} shows the globally
integrated kinetic energies in the 0.5 $M_\odot$ case, as a function
of time from the start of the simulation.  The evolution during the
first year is characterized by an initially linear (exponential)
growth of the convective instability, followed by a nonlinear
saturation phase, then by relaxation oscillations until it settles
down into a statistically steady state.  The longer term evolution
exhibits a growth in the differential rotation, which also reaches a
steady state.  It is clear from Figure \ref{fig_scalareng} that there
is little change during the last 5000 days of evolution.  The
results in Figure \ref{fig_fluxbalance} and presented below are shown
during this mature state of the systems.

\section{3D simulations} \label{sec_results}

The ending time of each of the 4 simulations ($t_f$) is listed in
table \ref{tab_results}.  In this section, we present the convective
properties and the global flows existing in each case.

\subsection{Properties of convection}

Figure \ref{fig_ss} shows the radial velocity on slices of constant
radius at slightly below the top of the domain and at the middle of
the domain, for the 0.5 (top), 0.7 (middle), and 1.1 (bottom)
$M_\odot$ cases.  Also shown are the slices of temperature at the
middle of the domain (right panels).  The temperature slices generally
appear less structured than the radial velocity slices because the
Prandtl number is less than unity (0.25), so that the thermal
difussivity ($\kappa$) is larger than the kinematic viscosity ($\nu$).


The convective patterns and characteristic eddy sizes are influenced
strongly by the density scale height, as well as by the contrast
between horizontal and vertical velocity \citep[see,
e.g.,][]{Bessolaz:2011p3250}.  Near the surface, where the density
scale height is smallest, the shell slices exhibit typical convective
patterns characterized by hot, broad/patchy upflows surrounded by
cool, narrow downflow lanes \citep[e.g.,][]{Cattaneo:1991p3746,
  Miesch:2008p3747}.  Deeper in the convection zone, the downflow
lanes often merge and generally lose their connectivity with respect
to the surface patterns, and the convection has a less patchy
appearance.  The near-surface shell slices (left panels of Fig.\
\ref{fig_ss}) are shown at a location where the unresolved eddy flux
dominates the other fluxes (see Fig.\ \ref{fig_fluxbalance}).  The
unresolved eddy flux is a spherically symmetric quantity, and it does
not affect the shape of the convective patterns, but it does influence
the level of (resolved) turbulence, since the unresolved eddy flux
reduces the convective driving.  However, at the location of the
near-surface shell slices shown, all cases are still turbulent, with a
significant convective (enthalpy) flux and Reynolds numbers in the
range of $10-30$.

In all of our cases, the convective patterns evolve on typical
timescales of a few weeks to a few months, with cells merging,
splitting, and disappearing.  The evolution of the convective motions
in the 1.1 $M_\odot$ case generally evolve on timescales of a factor
of a few times shorter than in the 0.5 $M_\odot$ case.  The convective
patterns are partly advected by the local shear and, depending on
whether the flow is prograde of retrograde, are tilted to the right or
the left near the equator.  At higher latitude, beyond the tangent
cylinder (an imaginary circle crossing the upper surface with a
cylindrical radius of $R_{\rm cz}$), the convective cells are less
aligned with the rotation axis and exhibit a more patchy behavior.

\begin{figure*}
\includegraphics[width=.5\linewidth]{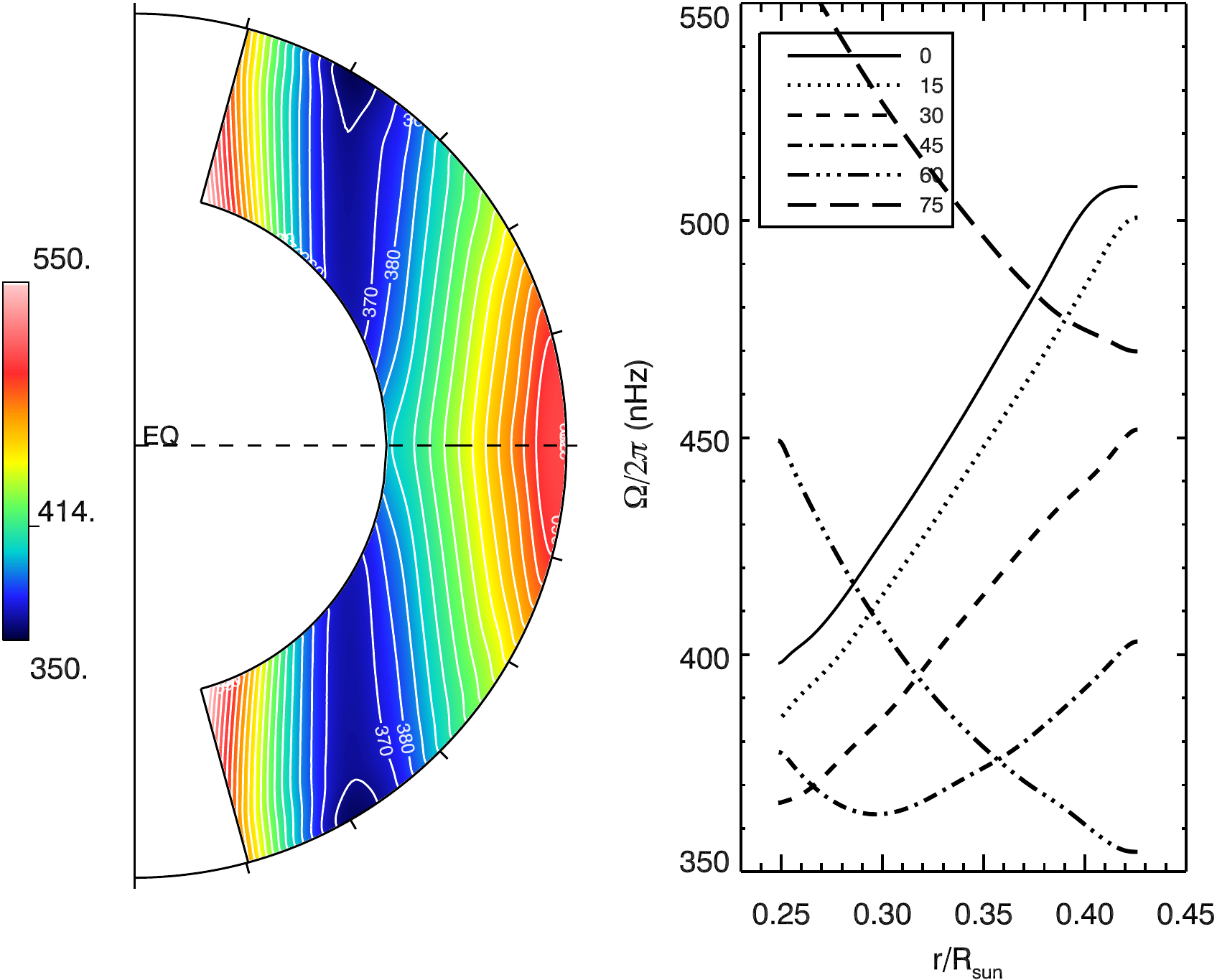}
\includegraphics[width=.5\linewidth]{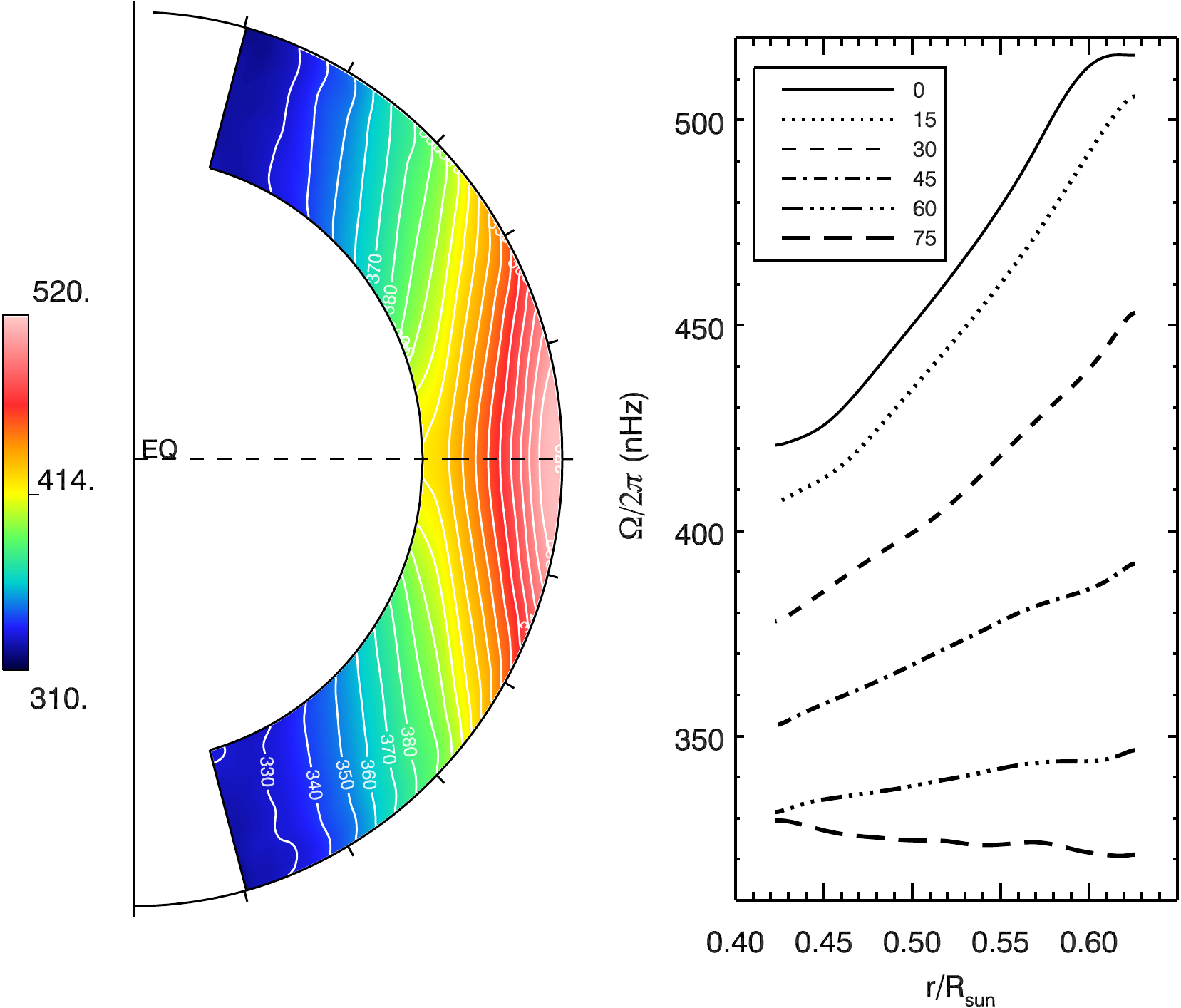}
\includegraphics[width=.5\linewidth]{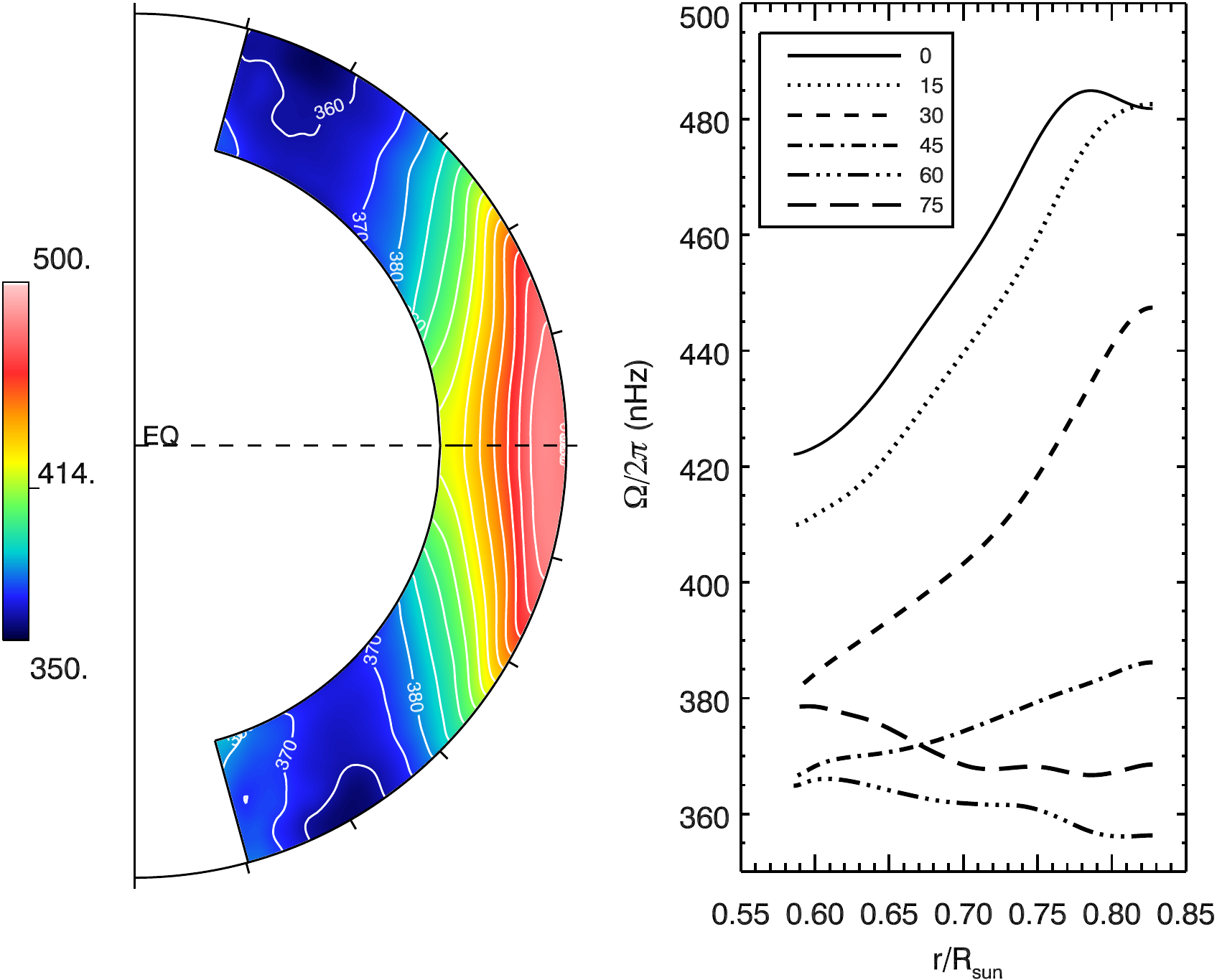}
\includegraphics[width=.5\linewidth]{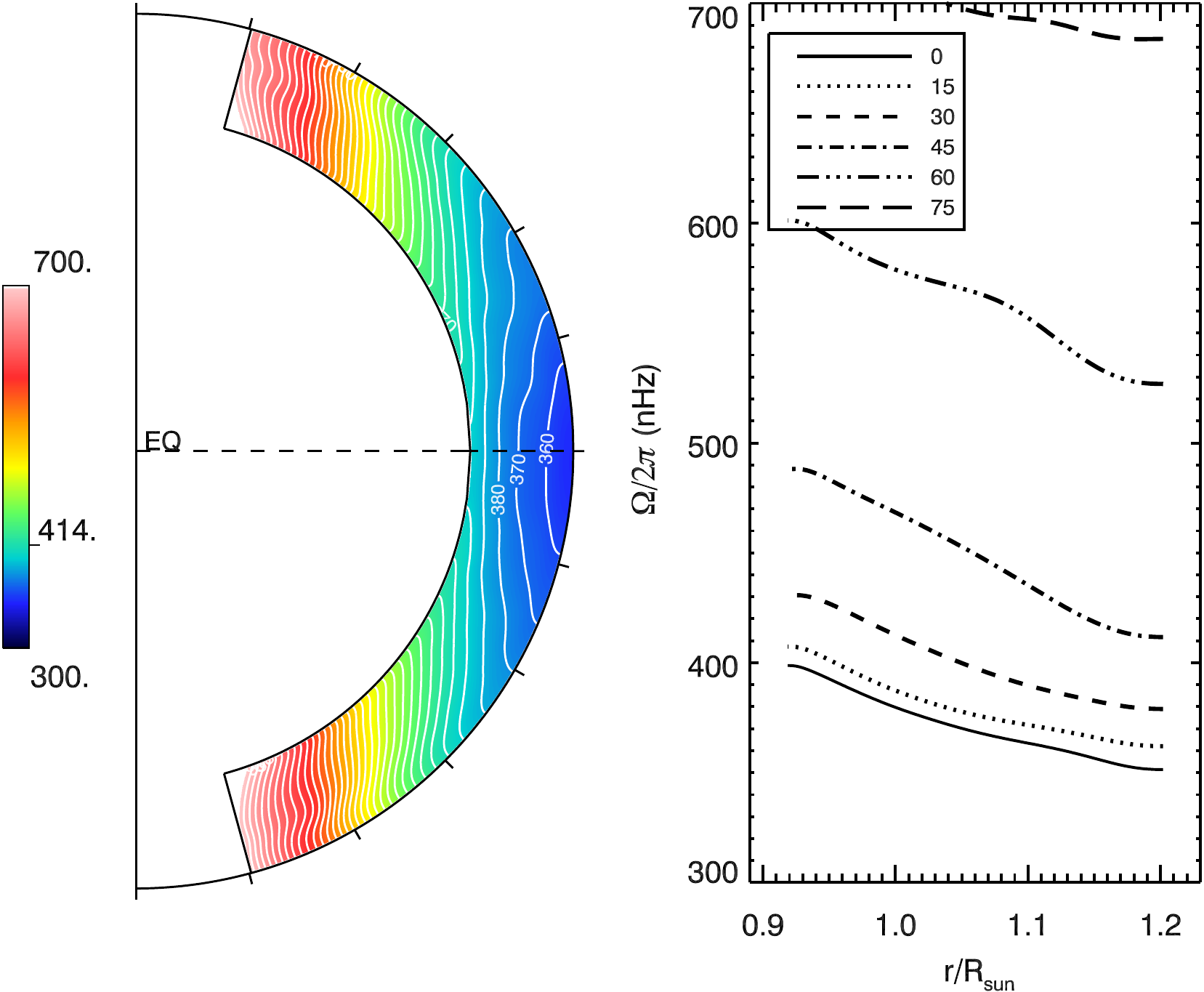}
\caption{Azimuthally and temporally averaged Rotation frequency in a
  meridional slice through the domain (contour/color plots; rotation
  axis is vertical) and shown along radial lines at different
  latitudes (line plots, latitude given in degrees).  Shown are the
  0.5 (top left), 0.7 (top right), 0.9 (bottom left), and 1.1 (bottom
  right) $M_\odot$ cases.}
\label{fig_diffrot}
\end{figure*}





As evident in the Figure, the 0.5 $M_\odot$ case exhibits convective
patterns near the equatorial region that are inhomogeneous, consisting
of vigorous convection over some range of longitudes and relatively
quiescent regions at other longitudes.  Such ``active nest''
convective patterns have been discussed by \citet{Ballot:2007p1132}
and \citet{Brown:2008p1119} and are likely related to our choice of
low Prandtl number.  The remaining 3 cases, exhibit vigorous
convection within the entire volume of the convection zone, as seen
for the 0.7 and 1.1 $M_\odot$ cases shown in the bottom two rows of
Figure \ref{fig_ss}.

Table \ref{tab_results} lists the rms velocities at the middle of the
computational domain for each case.  There is a clear trend of
increasing convective velocities with increasing stellar mass.  As
evident in the right panels of Figure \ref{fig_ss} and in the values
of the rms temperature ($\tilde T$) listed in table \ref{tab_results},
there is also a strong trend of increasing temperature contrast
between up- and downflows, with increasing mass.  This trend is
steeper if one considers the temperature fluctuation relative to the
mean temperature, since the mean background temperature in the 1.1
$M_\odot$ convection zone midlevel is $\sim$3 times colder than for
the 0.5 $M_\odot$ star (see \S \ref{sec_1d}).  Furthermore, the
average mass density in the convective envelope decreases with
increasing stellar mass (the midlevel density in the 1.1 $M_\odot$
case is 280 times lower than in the 0.5 $M_\odot$ case), which
decreases the efficiency of convective energy transport.  In spite of
the lower density, the convection in higher mass stars is able to
carry a larger luminosity than the lower mass stars, due to a
combination of larger stellar radii, higher convective velocities, and
higher temperature contrasts between up- and downflows.

Table \ref{tab_results} also lists the Reynolds, Rossby, and P\'eclet
numbers for each model.  The Reynolds (and P\'eclet) have a range of a
factor of a few, with no trend in mass.  These values primarily
reflect our choice of $\nu_{\rm top}$ for each case (tab.\
\ref{tab_parameters}) rather than any intrinsic property of the stars.
For our choices of $\nu_{\rm top}$, the 0.5 $M_\odot$ case has the
highest Reynolds number.  On the other hand, the Rossby numbers do not
depend strongly on our choice of diffusivities, and they show a strong
dependence on mass, primarily due to the trend in convective velocity
with mass, at a constant rotation period.  Finally, table
\ref{tab_results} lists a characteristic convective turnover timescale
for each case.  This also shows a strong trend with mass, primarily
due to the trend in convective velocities.

\subsection{Differential rotation and meridional circulation}

Figure \ref{fig_diffrot} shows the differential rotation in all 4
cases.  The 2D plots in the Figure exhibit isorotation contours that
are nearly aligned on cylinders in all cases.  This behavior is
typical for simulations such as these that have no enhanced latitudinal entropy
gradient present at the lower boundary of the convection zone
\citep{Miesch:2006p1251, Ballot:2007p1132}.

The angular rotation rate in the 0.5 $M_\odot$ case has a minimum
value at mid latitudes, a local maximum in the outer equatorial
region, and a global maximum near the pole.  A similarly ``banded''
differential rotation pattern is evident in a number of previously
published ASH simulations \citep[e.g.][]{Browning:2008p1139,
  Bessolaz:2011p3250}.  The differential rotation in the 0.7 and 0.9
$M_\odot$ cases are the most solar-like, with the fastest angular
rotation rate at the equator and slowest at higher latitudes.
However, the 1.1 $M_\odot$ case exhibits anti-solar rotation, where
the slowest angular velocity is at the equator.  A reversal of the
sense of differential rotation was also observed in the simulations of
\citet{Bessolaz:2011p3250} to be an effect of the thickness of the
convection zone.  In that study, the stellar structure and luminosity
was held fixed, while only the thickness of the convection zone was
varried.  The thickness of the convection zone in our 1.1 $M_\odot$
case (as a fraction of $R_*$) is comparable to the models of
\citet{Bessolaz:2011p3250} showing anti-solar rotation.  This would
suggest that the switch from solar to anti-solar may have more to do
with convection zone thickness than with stellar structural
properties.  At the same time, our choice of parameters may also have
influenced this outcome.  In particular, the 1.1 $M_\odot$ case has
the largest Rossby numbers (see tabs.\ \ref{tab_parameters} and
\ref{tab_results}), which means that the convection is less influenced
by rotation than in the other cases.

Each case exhibits a time-averaged, axisymmetric, latitudinal
temperature gradient.  For the 3 highest mass cases, this is
characterized by a monotonic change from equator to pole.  The 1.1
$M_\odot$ case has a hotter equator, while the 0.7 and 0.9 $M_\odot$
cases have hotter poles (as in the sun).  The 0.5 $M_\odot$ case has a
more complex temperature pattern with a maximum at mid latitudes,
while the latitudinal entropy gradient in all cases follows a more
monotonic behavior from equator to pole.  The second-to-last column of
table \ref{tab_results} shows the temporally and azimuthally averaged
temperature difference between the equator and a latitude of
60$^\circ$, at the base of the convection zone.  It is evident that
the temperature gradients are steaper for higher mass stars.  The more
complex temperature structure of the lowest mass case is consistent
with the banded differential rotation pattern, and the reversed
polarity of the temperature variation in the highest mass case is
consistent with anti-solar differential rotation.

The fractional differential rotation of each case, measured at the top
of the domain between latitudes 0 and 60 degrees, is listed in the
final column of table \ref{tab_results}.  These values of $\Delta
\Omega/\Omega_0$ in the table, show that the magnitude of the
differential rotation ranges from 30 to 42\% for all cases.


\begin{figure}
\includegraphics[width=1.05\linewidth]{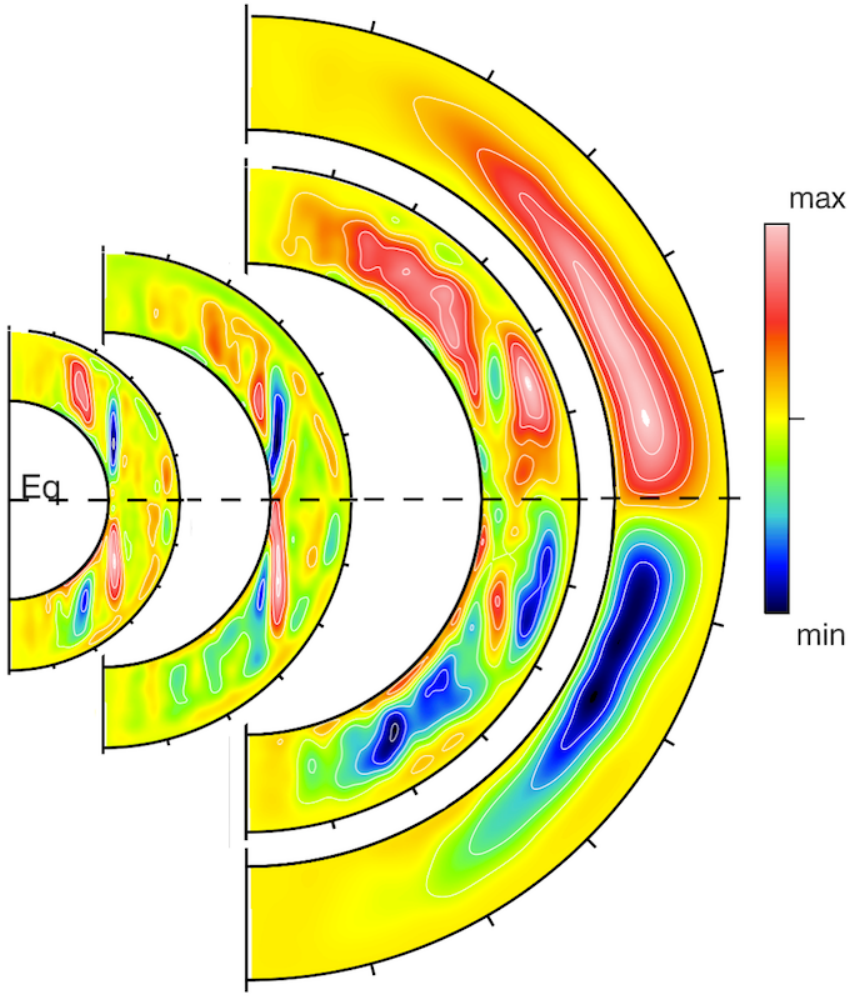} 
\caption{Contours of the stream function of the meridional flow for
  the 0.5, 0.7, 0.9, and 1.1 $M_\odot$ cases (left to right), where
  the physical sizes of the stars are shown to scale.  The contours
  follow streamlines in the meridional plane from temporally and
  azimuthally averaged data.  The colors indicate clockwise (blue) and
  counter-clockwise (red) circulation, and the stellar rotation axis
  is vertical.}
\label{fig_mcnested}
\end{figure}



Figure \ref{fig_mcnested} shows the meridional circulation for all 4
cases.  The plots are shown to scale, according to the physical size
of each of the stars.  For the 2 highest mass stars, the circulation
pattern is dominated by a global, unicellular flow pattern from
equator to pole near the surface and from pole to equator near the
base of the convection zone.  However, as evident in the Figure, there
is a trend for this flow to break up into a multi-cellular network of
circulation patterns roughly aligned on cylinders, for the lower mass
stars.  The tendancy for the meridional flow to breakup in this way
for increasing convection zone depth was also seen in the simulations
of \citet{Bessolaz:2011p3250}, suggesting that this effect is mostly
due to the size of the convection zone, rather than with other stellar
properties.

The last column of table \ref{tab_results} shows the characteristic
meridional circulation flow speed evaluated at the midlevel of the
domain for each case.  As with the convective velocities, there is a
clear trend of increasing meridional circulation flow speed with
increasing stellar mass.  A similar trend is also seen in 2D models
\citep[e.g.,][]{Kuker:2008p3594}.  Since the physical size of the
circulating patterns (evident in Fig.\ \ref{fig_mcnested}) also
increases strongly with mass, this suggests that the characteristic
timescale for these circulations is not strongly dependent on stellar
mass.  The values listed in table \ref{tab_results} also suggest that
the ratio of the meridional circulation speed to the convection speed
(e.g., $\tilde v_{\rm mc} / \tilde v$) range from 2\% for the 3 lowest
mass stars to 9\% for the 1.1 $M_\odot$ case.

\section{Summary \& discussion}
\label{sec_summary}

We have presented results from 3D dynamical simulations of the
convection zones of 4 sun-like stars, spanning a mass range from 0.5
to 1.1 $M_\odot$ and rotating at the solar rate.  As presented in
section \ref{sec_1d}, these stars cover a range in luminosities of a
factor of 40 and have convection zone masses ranging from 18\% to
1.1\% of a solar mass.  The physical sizes of the convection zones
increase slightly with mass (ranging from 19\% to 31\% of a solar
radius), although the fractional size decreases with mass (ranging
from 44\% to 24\% of a stellar radius).

We found that the convective velocities and temperature contrast
between up- and downflows is a strong function of mass.  This is due
to the strong luminosity dependence on mass, coupled with the fact
that higher mass stars have less dense convective envelopes and thus
require more vigorous convection.  As a consequence of the trend in
convective velocities, the Rossby number (for a constant $\Omega_0$)
increases and the convective turnover time decreases significantly
with increasing mass.

The convective pattern in the 0.5 $M_\odot$ case exhibits ``active
nests'' on a finite range of longitudes in the equatorial region.  The
presence of these active nests appears to be related to both the
Rossby and Prandtl numbers, in a sense that they have been seen in
simulations with low $R_o$ number and with $P_r=0.25$
\citep{Ballot:2007p1132, Brown:2008p1119} but not in cases with
$P_r=1.0$ \citep{Ballot:2007p1132, Bessolaz:2011p3250}.  This may be
due to the relative amplitude of the local shear, which tends to be
stronger for lower $P_r$.  Stronger shear disrupts convective motions
and can locally inhibit radial motions.

The angular rotation rate has a peak value at the surface and equator
for the 0.5, 0.7, and 0.9 $M_\odot$ cases.  However, our simulations
exhibit a differential rotation {\b that is} ``banded,'' with a
minimum at mid latitudes and another maximum at the poles, for the 0.5
$M_\odot$ case.  The 0.7 and 0.9 $M_\odot$ cases exhibit the
most solar-like differential rotation, with a maximum at the equator
and minimum near the poles, but the 1.1 $M_\odot$ case exhibits
anti-solar differential rotation.  The magnitude of the differential
rotation is comparable in all cases.

The 4 simulations presented here exhibit isorotation contours that are
nearly aligned on cylinders, approximately following the
Taylor-Proudman constraint.  From simulations of the solar convection
zone \citep[e.g.,][]{Miesch:2006p1251}, we know that the
Taylor-Proudman constraint can be broken by a thermal wind driven by
very small latitudinal gradients of temperature or entropy, with
differences of a few Kelvin sufficing to move cylindrical profiles to
radially-aligned rotation profiles.  Such gradients and thermal winds
do occur in the simulations presented here, but the presence of the
domain boundary at the base of the convection zone somewhat inhibits
this effect.  By moving the lower domain boundary deep into the
radiation zone, \citet{brun3ea11} demonstrated that the presence of a
dynamically self-consistent tachocline alleviates the effects of the
lower boundary on the convection zone dynamics and results in a
stronger thermal wind, isorotation contours more aligned to the radial
direction, and more realistic meridional circulations and convective
penetration.  In future work, similar models will be developed for the
stars in the present study.

The meridional circulation is dominated by a single, global
circulation pattern in each hemisphere for the most massive cases, but
breaks up into smaller scale patterns, aligned with the cylindrical
z-direction, for lower mass stars.

Future work with these models should explore cases that include, for
example, (a) a part of the radiation zone in the computational domain,
in order to determine how the presence of a tachocline affects the
convection zone properties, (b) the effects of different rotation
rates, in particular faster rotation, to understand how the properties
of these stars differ at younger ages, and (c) magnetic fields, in order to
address the magnetic dynamo properties of these stars.

\acknowledgements

SPM, ODC, and ASB were supported by the ERC through grant 207430
STARS2.  BPB is supported in part by NSF Astronomy and Astrophysics
postdoctoral fellowship AST 09-02004.  CMSO is supported by NSF grant
PHY 08-21899.




\end{document}